\documentclass[journal=jacsat,manuscript=article]{achemso}

\usepackage[version=3]{mhchem} % Formula subscripts using \ce{}

\usepackage{achemso}
\usepackage{times}
\usepackage{enumitem}
\usepackage{color, soul}
\usepackage{graphicx}
\graphicspath{ {Figures/} }
\usepackage[table]{xcolor}
\usepackage[labelfont=bf]{caption}
\usepackage{amsmath} 
\usepackage{amssymb} 
\usepackage{bm}
\usepackage{float}
\title{Magnetic Proximity induced efficient charge-to-spin conversion in large area PtSe$_{2}$/Ni$_{80}$Fe$_{20}$ heterostructures}

\author{Richa Mudgal}
\affiliation[Indian Institute of Technology Delhi]
{Department of Physics, Indian Institute of Technology Delhi, Hauz Khas, New Delhi-110016, India}
\author{Alka Jakhar}
\affiliation[Indian Institute of Technology Delhi]
{Center for Applied Research in Electronics, Indian Institute of Technology Delhi, Hauz Khas, New Delhi-110016, India}
\author{Pankhuri Gupta}
\affiliation[Indian Institute of Technology Delhi]
{Department of Physics, Indian Institute of Technology Delhi, Hauz Khas, New Delhi-110016, India}
\author{Ram Singh Yadav}
\affiliation[Indian Institute of Technology Delhi]
{Department of Physics, Indian Institute of Technology Delhi, Hauz Khas, New Delhi-110016, India}
\author{B. Biswal}
\affiliation[Indian Institute of Technology Madras]
{Condensed Matter Theory and Computational Lab, Department of Physics, Indian Institute of Technology Madras, Chennai 600036, India}
\alsoaffiliation[Indian Institute of Technology Madras]
{Center for Atomistic Modelling and Materials Design, Indian Institute of Technology Madras, Chennai 600036, India}
\author{P. Sahu}
\affiliation[Indian Institute of Technology Madras]
{Condensed Matter Theory and Computational Lab, Department of Physics, Indian Institute of Technology Madras, Chennai 600036, India}
\alsoaffiliation
{Department of Physics \& Astronomy, University of Missouri, Columbia, MO 65211, USA}
\author{Himanshu Bangar}
\affiliation[Indian Institute of Technology Delhi]
{Department of Physics, Indian Institute of Technology Delhi, Hauz Khas, New Delhi-110016, India}
\author{Akash Kumar}
\affiliation[Indian Institute of Technology Delhi]
{Department of Physics, Indian Institute of Technology Delhi, Hauz Khas, New Delhi-110016, India}
\alsoaffiliation[University of Gothenburg]
{Department of Physics, University of Gothenburg, Gothenburg-412 96, Sweden}
\author{Niru Chowdhury}
\affiliation[Indian Institute of Technology Delhi]
{Department of Physics, Indian Institute of Technology Delhi, Hauz Khas, New Delhi-110016, India}
\author{Biswarup Satpati}
\affiliation[Saha Institute of Nuclear Physics]{Surface Physics \& Material Science Division, Saha Institute of Nuclear Physics, A CI of Homi Bhabha National Institute, 1/AF Bidhannagar, Kolkata 700064, India}
\author{B. R. K. Nanda}
\affiliation[Indian Institute of Technology Madras]
{Condensed Matter Theory and Computational Lab, Department of Physics, Indian Institute of Technology Madras, Chennai 600036, India}
\alsoaffiliation[Indian Institute of Technology Madras]
{Center for Atomistic Modelling and Materials Design, Indian Institute of Technology Madras, Chennai 600036, India}
\author{S. Satpathy}
\affiliation[Indian Institute of Technology Madras]
{Condensed Matter Theory and Computational Lab, Department of Physics, Indian Institute of Technology Madras, Chennai 600036, India}
\alsoaffiliation
{Department of Physics \& Astronomy, University of Missouri, Columbia, MO 65211, USA}
\alsoaffiliation[Indian Institute of Technology Madras]
{Center for Atomistic Modelling and Materials Design, Indian Institute of Technology Madras, Chennai 600036, India}
\author{Samaresh Das}
\affiliation[Indian Institute of Technology Delhi]
{Center for Applied Research in Electronics, Indian Institute of Technology Delhi, Hauz Khas, New Delhi-110016, India}
\author{P. K. Muduli}
\email{muduli@physics.iitd.ac.in}
\affiliation[Indian Institute of Technology Delhi]
{Department of Physics, Indian Institute of Technology Delhi, Hauz Khas, New Delhi-110016, India}
\begin{document}
\newpage
\begin{abstract}
%  provide an energy-efficient method for manipulating the magnetization dynamics of ferromagnets. Non-magnetic materials with large charge-to-spin conversion efficiency are highly desired for SOT applications. 
 As a topological Dirac semimetal with controllable spin-orbit coupling and conductivity, PtSe$_2$, a transition-metal dichalcogenide, is a promising material for several applications from optoelectric to sensors. However, its potential for spintronics applications is yet to be explored. In this work, we demonstrate that PtSe$_{2}$/Ni$_{80}$Fe$_{20}$ heterostructure can generate a large damping-like current-induced spin-orbit torques (SOT), despite the absence of spin-splitting in bulk PtSe$_{2}$. The efficiency of charge-to-spin conversion is found to be $(-0.1 \pm 0.02)$~nm$^{-1}$ in PtSe$_{2}$/Ni$_{80}$Fe$_{20}$, which is three times that of the control sample, Ni$_{80}$Fe$_{20}$/Pt. Our band structure calculations show that the SOT due to the PtSe$_2$ arises from an unexpectedly large spin splitting in the interfacial region of PtSe$_2$ introduced by the proximity magnetic field of the Ni$_{80}$Fe$_{20}$ layer. Our results open up the possibilities of using large-area PtSe$_{2}$ for energy-efficient nanoscale devices by utilizing the proximity-induced SOT.
 
\end{abstract}

%\textbf{Keywords:} Transition metal dichalcogenides (TMDs); Spin-orbit torque (SOT), Charge to spin
%conversion; Spin
%torque ferromagnetic resonance; Proximity effect

%\section{\label{sec:level1}Introduction}
%\vspace{-1mm}

The next generation of spintronic memories can be realized using spin-orbit torques (SOT), an energy-efficient and faster way to manipulate magnetization.~\cite{manchon2019current}
In ferromagnet (FM)/heavy metal (HM) bilayers, SOT describes the current-induced torque exerted on the FM layer. The flow of a charge current in an FM/HM system can generate a spin current, transverse to the direction of charge current flow, by mechanisms such as the spin Hall effect,~\cite{dyakonov1971current} or Rashba-Edelstein.~\cite{bychkov1984properties} Typically heavy metals, such as Pt,~\cite{liu2011spin,skowronski2019determination} W,~\cite{pai2012spin,bansal2018large,skowronski2019determination} and Ta~\cite{liu2012spin,kumar2018large} are used to generate spin currents.
Significant efforts have been reported to improve the energy efficiency of HM-based SOT devices.~\cite{demasius2016enhanced,kumar2021large,behera2022energy} One method to achieve a higher SOT efficiency is by replacing the HM layer with alternative materials that have large charge-to-spin conversion efficiency and high conductivity.
Transition metal dichalcogenides (TMDs) are potential replacements of HM due to various properties such as tunable conductivity,~\cite{chen2014thickness,behera2022tunable} high spin–orbit coupling,~\cite{zhu2011giant,bansal2019extrinsic,bangar2022large} the presence of Dzyaloshinskii Moriya-interaction,~\cite{kumar2020direct} tunable band structure,~\cite{kang2017universal,nugera2022bandgap} spin-layer locking~\cite{yao2017direct} and long spin-life time.~\cite{liang2017electrical} The use of TMDs in SOT devices has shown a number of advantages, \textit{e.g.,} the ability to control SOT using the crystal symmetry~\cite{macneill2017control, kao2022deterministic,zhao2020unconventional} and electric field.~\cite{lv2018electric}

%Despite these advantages, many TMDs have low conductivity, which limits their applications for SOT devices. However, platinum-based TMDs are very promising for SOT devices due to their high electrical conductivity compared to other TMDs.~\cite{zhao2017high} 

The Pt-based TMDs offer advantages such as high air stability,\cite{zhao2017high}, and large spin-orbit coupling.~\cite{husain2020emergence} However, SOT studies on Pt-based TMDs have been limited to PtTe$_2$ for which a recent work showed a large damping-like torque and SOT switching.~\cite{xu2020high} PtSe$_2$ is another Pt-based TMD offering high and tunable spin-orbit coupling.~\cite{husain2020emergence} It is also a type-II Dirac semi-metal, offering several topologically protected phenomena potentially useful for spintronic applications.~\cite{zhang2017experimental} In addition, the band structure of PtSe$_2$ exhibits topological surface states,~\cite{bahramy2018ubiquitous} which can be exploited for topological spintronic devices. It has been demonstrated that the monolayer PtSe$_2$ film possesses spin-layer locking~\cite{yao2017direct} with many interesting electronic properties.~\cite{wang2015monolayer}
These intriguing properties suggest that PtSe$_{2}$ might be a promising candidate for investigating the SOT phenomenon. However, unlike other TMDs~\cite{bangar2022large}, the band structure of PtSe$_2$ does not show spin splitting. In addition, the type II Dirac cone appears much below the Fermi energy in the bulk PtSe$_2$.~\cite{zhang2017experimental} %Recent studies show that the heterostructures of TMD and FM can exhibit proximity-induced enhanced SOT.~\cite{li2023enhancement,dolui2020spin} Such a possibility has not yet been investigated in PtSe$_2$. 

In this work, we investigate SOT in PtSe$_2$/Ni$_{80}$Fe$_{20}$(or NiFe)/Pt heterostructures using the spin torque ferromagnetic resonance (STFMR) technique. We found a large increase in the effective damping constant (164\%) in PtSe$_2$/NiFe/Pt compared to the NiFe/Pt control sample. Using angle-resolved STFMR measurements, we demonstrated that only conventional SOTs are generated in PtSe$_2$/NiFe/Pt, with spin polarization along the \textit{y-}axis for a charge current flowing along the \textit{x-}direction, which is in accordance with theoretical predictions.~\cite{liu2020two} From the line shape analysis, we determine an enhanced effective damping-like torque due to PtSe$_2$. We found a maximum interfacial charge-to-spin conversion efficiency of $-0.1 \pm 0.02$~nm$^{-1}$ due to PtSe$_2$. Our band structure calculation of PtSe$_2$ that includes the proximity of the FM layer shows that NiFe introduces: (a) charge transfer from NiFe to PtSe$_2$, (b) a Rashba SOC due to the surface symmetry breaking, and (c) a proximity magnetic field induced spin-splitting that could lead to enhanced spin Hall conductivity (SHC) in the interface region. The enhanced SOT due to PtSe$_2$ is qualitatively explained using the substantial spin splitting introduced by the proximity magnetic field.
\begin{figure}
    \centering
    \includegraphics[width=17cm]{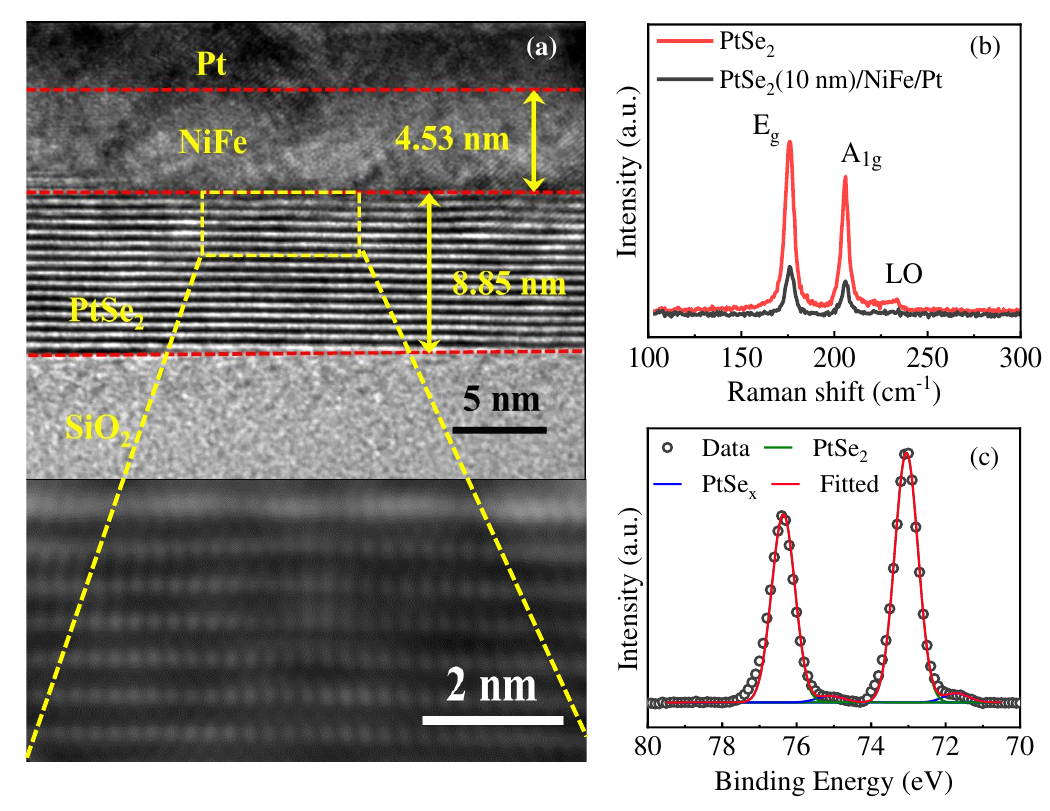}
    \caption{(a) cross-sectional HRTEM image of PtSe$_2$/NiFe/Pt stack. The interface is indicated by the red dashed lines. The magnified view of the square region is shown in the bottom part of the figure, showing the formation of high-quality PtSe$_2$ lattices (with 1T structure). (b) Raman spectra of PtSe$_{2}$ thin film (red curve) and that of the device after deposition of NiFe/Pt (black curve). The spectra show two strong peaks labeled as E$_\mathrm{g}$ and A$_\mathrm{1g}$ and one shoulder peak labeled as longitudinal optical (LO) mode. (c) X-ray photoelectron spectra of Pt-4f core level. Spectra were fitted with four peaks corresponding to PtSe$_2$ and PtSe$_x$. Splitting between doublets of Pt-4f peaks is found to be 3.3~eV.}
    \label{fig:crystal}
\end{figure}

 We used a thermally assisted conversion process to grow large-area PtSe$_{2}$ thin films.~\cite{jakhar2020optically}  
 HRTEM and XRD (section S1 in supplementary) measurements on PtSe$_2$ show the growth of crystalline films.   Figure~\ref{fig:crystal} (a) shows the cross-sectional HRTEM image of PtSe$_2$/NiFe/Pt stack. 
 %The HRTEM measurements were performed on an FEI, TECNAI G2 F30, ST microscope with an accelerating voltage 300 kV. 
 The image shows that the whole Pt layer gets selenized, and a high-quality PtSe$_2$ lattice with a 1T structure is formed. The thickness of PtSe$_2$ is found $\approx$ 9 nm, and the separation between two consecutive layers is $\approx$ 0.5 nm. Each layer corresponds to a monolayer of PtSe$_2$, which contains one Pt layer sandwiched between two Se layers. A magnified view of the square region is shown at the bottom of the figure.  %, which shows the atoms in the PtSe$_2$ layer.  
 Based on HRTEM and AFM measurements (Section S1 in supplementary), we determine the thickness of PtSe$_{2}$ after selenization of 3~nm thick Pt to be (10 $\pm$ 1.0)~nm. This increase in the thickness of Pt due to its selenization is well-known in the literature,~\cite{yim2018electrical} and it further confirms the complete selenization of the Pt layer.
 Figure~\ref{fig:crystal} (b) shows the Raman spectra of PtSe$_2$, measured with a laser of wavelength 532 nm. The spectrum contains two distinct peaks that corresponds to PtSe$_2$: E$_\mathrm{g}$ at 176 cm$^{-1}$ and A$_\mathrm{1g}$ at 206 cm$^{-1}$. Here, the E$_\mathrm{g}$ corresponds to the in-plane vibrations of Se atoms moving away from each other while the A$_\mathrm{1g}$ corresponds to out-of-plane vibrations of Se atoms moving in the opposite directions.~\cite{yan2017high} A small peak labeled as longitudinal optical (LO) mode is also observed at 233 cm$^{-1}$, which is a combination of in-plane and out-of-plane vibrations of Pt and Se atoms.~\cite{yan2017high, gulo2020temperature} The position and intensity ratio of A$_\mathrm{1g}$ and E$_\mathrm{g}$ peaks confirm the presence of PtSe$_2$ in bulk form.~\cite{yan2017high}
Similar spectra were obtained at multiple locations on the sample with a ratio of the intensity of A$_\mathrm{1g}$ and E$_\mathrm{g}$ peaks nearly identical (0.75 $\pm$ 0.02), confirming the formation of a large-area homogeneous thin film of PtSe$_{2}$. The chemical composition of PtSe$_2$ was verified further by X-ray photoelectron spectroscopy (XPS). Figure~\ref{fig:crystal} (c) shows the Pt-4f core-level spectra. The binding energies of the different core levels have been calibrated using the C-1s peak and the core-level
spectra have been best fitted with Gaussian profiles. Peaks corresponding to IV valency of Pt are observed at 76.3 and 73 eV, while peaks centered at 75.1 and 71.8 eV show the presence of PtSe$_x$ phase in a small amount. The splitting between doublet for each phase is found to be 3.3 eV, consistent with literature~\cite{su2018phase}. 

Subsequently, we fabricated microstrip ($100\times 40~\mu$m$^2$) devices for STFMR measurements using maskless lithography and lift-off method in two levels. To confirm the presence of PtSe$_{2}$ after deposition of NiFe/Pt and device fabrication, we perform Raman measurements [Fig.~\ref{fig:crystal} (b)] by focusing the laser on the microstrip. The intensity ratio of A$_\mathrm{1g}$ and E$_\mathrm{g}$ was found to be similar compared to pure PtSe$_{2}$, indicating that the quality of PtSe$_{2}$ was not affected by the deposition of NiFe/Pt and our fabrication steps. The overall reduction in the intensity of the peaks is due to the finite thickness of the top metal layers.

%\section{\label{sec:level3}Results and Discussion}
%\vspace{-1mm}

\begin{figure}[ht]
    \centering
    \includegraphics[width=17cm]{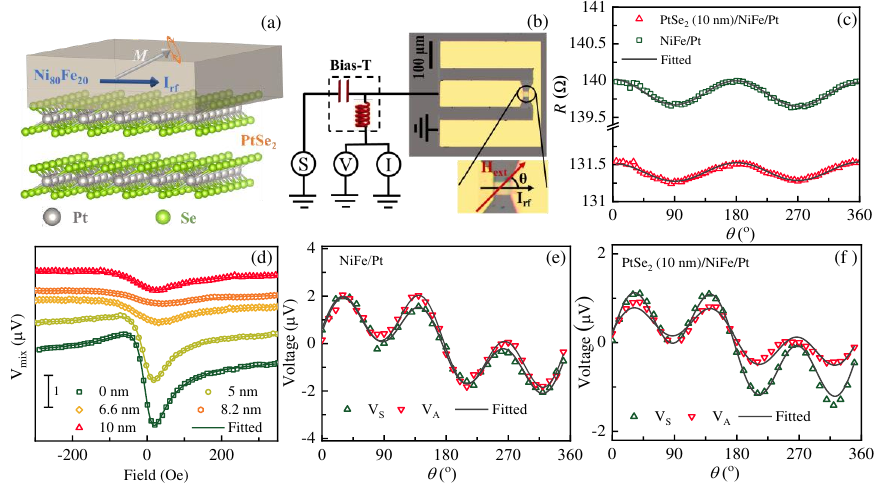}
    \caption{(a) Schematic of sample stack showing the direction of RF current and the precession of magnetization. (b) Schematic of STFMR set-up consisting of an optical image of the device. The directions of the RF current and applied field are also shown in the enlarged image. (c) Anisotropic magnetoresistance of PtSe$_2$ (10 nm)/NiFe (5 nm)/Pt (3 nm) to gather with control sample NiFe (5 nm)/Pt (3 nm).  (d) Example of STFMR spectra measured at 5~GHz and $\theta=50^{\circ}$ for PtSe$_2$(t)/NiFe/Pt for varying thickness of PtSe$_2$ layer. The plots are shifted for clarity. Angular dependence of $V_\mathrm{S}$ and $V_\mathrm{A}$ for (e) NiFe/Pt (f) PtSe$_2$ (10 nm)/NiFe/Pt, measured at 5 GHz. In all the plots, symbols represent the measured data, while solid lines represent fits as described in the main text.}
    \label{fig:set up}
\end{figure}

\begin{figure}
 \centering
 \includegraphics[width=17cm]{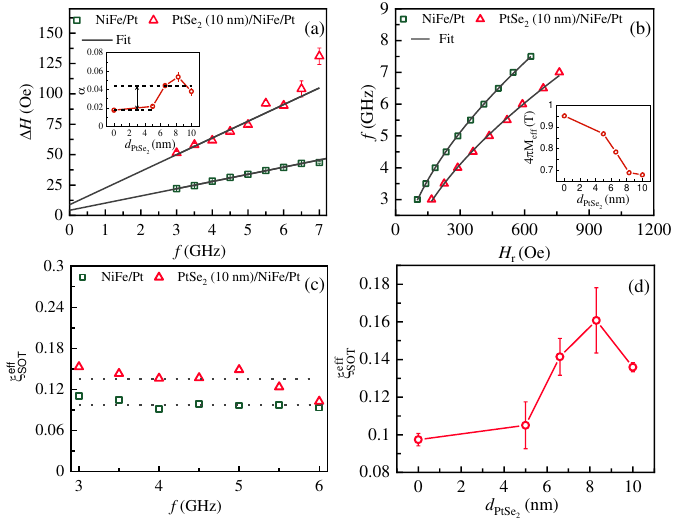}
 \caption{ (a) The linewidth versus frequency for NiFe/Pt and PtSe$_2$(10~nm)/NiFe/Pt. The inset shows a variation of the Gilbert damping parameter~($\alpha$) with the thickness of PtSe$_2$ ($d_\mathrm{PtSe_2}$) in the PtSe$_2$/NiFe/Pt stack. The dashed line in the inset shows the average value of $\alpha$ for $d_\mathrm{PtSe_2}\geq 6.5$~nm (b) Resonance field versus frequency for NiFe/Pt and PtSe$_2$(10~nm)/NiFe/Pt. The inset shows the variation of effective magnetization (M$_\mathrm{eff}$) with $d_\mathrm{PtSe_2}$. (c) Variation of effective SOT efficiency with frequency for PtSe$_2$(10)/NiFe/Pt and NiFe/Pt. The dotted lines represent the average values. (d) Variation of effective SOT efficiency, $\xi_\mathrm{SOT}^\mathrm{eff}$ with $d_\mathrm{PtSe_2}$. Each data point represents the average of three devices.}
 \label{fig:3}
\end{figure}

Figure~\ref{fig:set up} (a) depicts a schematic representation of the sample stack used for STFMR experiments, while Fig.~\ref{fig:set up} (b) shows the schematic of the STFMR circuit.~\cite {kumar2021large} An amplitude-modulated radio frequency (RF) current, $I_\mathrm{RF}$ is applied to the device through the RF port of a bias-tee. The DC port of the bias-tee is used for both measuring the rectified mixing voltage using a lock-in amplifier and applying DC current. We vary the in-plane field angle, $\theta$ with respect to RF current, as defined in the zoomed-in image of Fig.~\ref{fig:set up} (b).
Figure~\ref{fig:set up} (c) shows the measured anisotropic magnetoresistance (AMR)
of PtSe$_2$ (10 nm)/NiFe (5 nm)/Pt (3 nm) to gather with control sample NiFe (5 nm)/Pt (3 nm). The solid lines are calculated AMR curve with $R = R_{\perp} + (R_{\parallel} - R_{\perp}) \cos^{2}\theta$, where $R_{\perp}$ denotes the resistance for $\theta = 90^{\circ}$ and $R_{\parallel}$ denotes the resistance for $\theta=0^{\circ}$. The percentage of AMR, \textit{i.e.,} $(R_{\parallel} - R_{\perp})/R_{\perp}$ is found to be 0.19\% for PtSe$_2$/NiFe/Pt and 0.24\% for NiFe/Pt. % The value of resistance of the two devices are found very similar (within device to device variation), indicating a high resistivity of PtSe$_2$. 
Using a parallel resistor model, we estimate that the resistance of PtSe$_2$ to be $\approx2~k\Omega$ and the resistivity to be $\approx1955~\mu\Omega$~cm. %For both cases, resistance follows behavior , which varies with $\theta$ following cos$^2\theta$ dependence. AMR confirms the presence of magnetic layer in the stack. 
Figure~\ref{fig:set up} (d) shows the STFMR spectra for PtSe$_2$/NiFe/Pt to gather with the control sample NiFe/Pt measured at 5~GHz with a magnetic field applied at an angle of $\theta =~50^{\circ}$ for various thickness of PtSe$_2$. The data for other frequencies can be found in supplementary Sec.~S2. The lineshape of STFMR spectra for PtSe$_2$/NiFe/Pt is found to be more symmetric compared to the control sample. The measured STFMR spectra can be fitted using the following equation:~\cite{liu2011spin,kumar2021large}
\begin{equation}\label{Vmix}
 V_{\rm{mix}}=V_{\rm S}\frac{\Delta H^2}{\Delta H^2+(H-H_\mathrm{r})^2}+V_{\rm A}\frac{\Delta H({H-H_\mathrm{r}})}{\Delta H^2+(H-H_\mathrm{r})^2}
\end{equation}

where, $H$ is the applied magnetic field, $\Delta H$ is the linewidth, %(defined as full width at half maxima), 
$H_\mathrm{r}$ is the resonance field, % $R$ is the device resistance, and . 
and $V_{\rm S}$ and $V_{\rm A}$ are the amplitudes of the symmetric and antisymmetric voltage components, respectively. %which are proportional to the in-plane damping-like torque and the out-of-plane torques, respectively. 
By fitting the measured $V_{\rm mix}$ with above equation, we determine ${\Delta H}$, $H_\mathrm{r}$, $V_{\rm S}$ and $V_{\rm A}$. First, we performed angular-dependent STFMR measurements to determine the spin polarization of the interface-generated spin current. The angular dependence of $V_{\rm S}$ and $V_{\rm A}$ is depicted in Fig.~\ref{fig:set up} (e) and (f) for NiFe/Pt and PtSe$_2$( 10 nm)/NiFe/Pt respectively, which follows the conventional $\sin2\theta\cos\theta$ dependence (solid lines), showing that the spin-polarization is along $y-$axis, for a charge current along the $x-$axis, which is similar to Pt and other heavy metals.~\cite{liu2011spin,macneill2017control}. Next, we determine the Gilbert damping constant, $\alpha$ by fitting the ${\Delta H}$ versus frequency data [Fig.~\ref{fig:3} (a)] using:\cite{liu2011spin,kumar2021large}

\begin{equation}\label{LW}
 \Delta H=\frac{2\pi\alpha}{\gamma}f+\Delta H_0
\end{equation}

 Here, $\gamma$ is the gyromagnetic ratio, and $\Delta H_0$ is inhomogeneous linewidth which depends on magnetic inhomogeneity. The extracted value of $\alpha$ was plotted with the thickness of PtSe$_2$ ($d_\mathrm{PtSe_2}$) in the \textit{inset} of Fig.~\ref{fig:3} (a). 
 %The damping constant $\alpha$ is found to be constant for PtSe$_2 \approx 5nm$ with respect to control sample, while it is increased for higher thickness. 
 The value of $\alpha$ increases and becomes nearly constant for $d_\mathrm{PtSe_2}\geq 6.5$~nm. The average value of $\alpha$ for $d_\mathrm{PtSe_2}\geq 6.5$~nm is (0.045 $\pm$ 0.007), which is 165\% higher than the control sample. From this, we determine spin mixing conductance to be $\approx 7.6\times10^{19}\textrm{m}^{-2}$, which is higher than other TMD/NiFe systems.~\cite{bangar2022large, bansal2019extrinsic} The inhomogeneous linewidth, $\Delta H_0$ is found to be similar for all the devices $\approx$ (5 $\pm$ 3)~Oe,  indicating no significant degradation in the quality of magnetic layer (NiFe) grown on PtSe$_2$. Furthermore, the linear behavior of linewidth with frequency indicates the absence of two-magnon scattering.~\cite{arias1999extrinsic}
 We calculate $M_{\mathrm{eff}}$ for different stacks by fitting the behavior of the frequency \textit{versus} resonance field [Fig.~\ref{fig:3} (b)] with the Kittel formula: 
 
\begin{equation}\label{kittel}
 f=\frac{\gamma}{2\pi}\sqrt{(H_\mathrm{r}+H_\mathrm{k})(H_\mathrm{r}+H_\mathrm{k}+4\pi M_{\mathrm{eff}})}
\end{equation}
Here, $H_\mathrm{k}$ and $M_{\mathrm{eff}}$ are the anisotropy field and effective magnetization of the FM layer, respectively. By fitting the experimental data with the above equation, we found $4\pi M_\mathrm{eff}$ decreases with the thickness of PtSe$_2$ as shown in the \textit{inset} of Fig.~\ref{fig:3}(b). The effective magnetization $4\pi$$M_\mathrm{eff}$ is given by $4\pi M_\mathrm{eff}=4\pi M_s-2K_s/M_s t_{\rm{NiFe}}$, where $K_s$ is the surface anisotropy constant, $t_\mathrm{NiFe}$ is the thickness of the FM layer, and $M_\mathrm{s}$ is the saturation magnetization. Since $K_\mathrm{s}$ is proportional to interfacial spin-orbit coupling, we believe the decrease of $M_{\rm {eff}}$ for PtSe$_2$/NiFe/Pt is caused by the increase in interfacial spin-orbit coupling due to \textit{d-d} hybridization, similar to previous reports for other TMD/FM systems.~\cite{wu2020tuning, jamilpanah2020interfacial}

%Next, we quantify the SOT in our system  by lineshape as linewidth method. We discussed in details that why the lineshape method is not well suited for our case.
 As stated previously, PtSe$_2$ has a resistance greater than $2~k\Omega$; therefore, the current flowing through the PtSe$_2$ layer is only 6\% of the total current. Consequently, we anticipate SOT arising from (i) the interface between PtSe$_2$ and NiFe and (ii) the spin Hall effect of the Pt capping. Here, we neglect self-induced torque in NiFe, which is normally one order of magnitude lower.~\cite{fu2022observation} In order to quantify charge-to-spin conversion due to PtSe$_2$, we have used line shape analysis. In this method, the SOT efficiency, $\xi_\mathrm{SOT}$ is expressed as:~\cite{liu2011spin} 
\begin{equation} \label{SHA}
    {\xi_\mathrm{SOT}=\frac{V_\mathrm{S}}{V_\mathrm{A}}\frac{e\mu_0M_\mathrm{s}t_\mathrm{NiFe} d}{\hbar}\sqrt{1+\frac{4\pi M_{\mathrm{eff}}}{H_\mathrm{r}}}}
\end{equation} 
Here, $e$ is the electronic charge, and $\hbar$ is the reduced Planck's constant.  Here, $d$ is the thickness of the non-magnetic layer. The above expression is valid when the field-like torque is negligible. In our case, we determined field-like torque from the anti-symmetric component of $V_\mathrm{mix}$ as discussed in the supplementary Sec.~S4. We found field-like torque to be negligible for both NiFe/Pt and PtSe$_2$/NiFe/Pt. However, the above equation is normally used for the case of FM/NM bilayer samples. For the trilayer system of PtSe$_2$/NiFe/Pt, the lineshape method represents an \textit{effective} spin-orbit torque efficiency, $\xi_\mathrm{SOT}^\mathrm{eff}$.~\cite{liu2019strain} Since the conductivity of PtSe$_2$, is much lower than that of Pt, we assume that the Oersted field is only due to the Pt layer. With these assumptions, we determine $\xi_\mathrm{SOT}^\mathrm{eff}$, which includes the contribution of the top Pt layer and the interfacial contribution of PtSe$_2$. 
For this case, we use  the thickness of Pt for $d$ in Eq.~\ref{SHA}.~\cite{liu2019strain}. In Fig~\ref{fig:3} (c), the value of $\xi_\mathrm{SOT}^\mathrm{eff}$ has been plotted with frequency. The average value of $\xi_\mathrm{SOT}^\mathrm{eff}$ was further calculated and plotted in Fig.~\ref{fig:3} (d) for different thicknesses of PtSe$_2$. The value of $\xi_\mathrm{SOT}^\mathrm{eff}$ for $d_\mathrm{PtSe_2}$ $\approx$5~nm, was found to be 
 very close to the control sample, but it increases upto $(50\pm 13)\%$ for higher thickness. The value of $\xi_\mathrm{SOT}^\mathrm{eff}$ is constant with $d_\mathrm{PtSe_2}$ $>5$~nm. We also performed dc Planar Hall effect measurements by fabricating Hall bars. These measurements also show a similar increase of damping-like torque as discussed in supplementary material Sec.~S5. The increase of $\xi_\mathrm{SOT}^\mathrm{eff}$ due to the presence of PtSe$_2$ indicates that the sign of spin current due to PtSe$_2$ is opposite to that of the Pt layer. In order to extract the contribution of PtSe$_2$, we use the symmetric and asymmetric part of the $V_\mathrm{mix}$ of the control sample (see supplementary Sec.~S3), which also eliminates SOT from the Pt cap layer [mechanisms (ii) mentioned above]. Considering the interfacial effect as the origin of SOT, we extract the interfacial charge-to-spin conversion ($q_\mathrm{ICS}$) following Kondou \textit{et al}.~\cite{kondou2016fermi} Here, $q_\mathrm{ICS}$ is defined as 3D spin current ($J_\mathrm{s}$) per unit 2D charge current ($j_\mathrm{c}$). The maximum interfacial charge-to-spin conversion due to PtSe$_2$ ($q_\mathrm{ICS}^{PtSe_2}$) is determined to be $-0.1$~nm$^{-1}$. For the control samples, the interfacial charge-to-spin conversion is $0.03$~nm$^{-1}$. Thus the magnitude of interfacial charge-to-spin conversion is nearly three times larger compared to Pt. $q_\mathrm{ICS}$ of PtSe$_2$ is also higher compared to
 that of PtTe$_2$.~\cite{xu2020high}. In this reference, the authors reported spin Hall angle/damping-like torque. Since we report an interfacial damping-like SOT, the efficiency of   PtTe$_2$-work needs to be divided by the thickness of PtTe$_2$ for comparison with our work. Accordingly, the $q_\mathrm{ICS}$ of PtTe$_2$-work is $0.003-0.033$~nm$^{-1}$, which is lower than our study. $q_\mathrm{ICS}$ value for PtSe$_2$ is also found to be higher compared to W, ($q_\mathrm{ICS}=0.06$~nm$^{-1}$)~\cite{wu2019room} Bi$_2$Te$_3$,($q_\mathrm{ICS}=0.05$~nm$^{-1}$) ~\cite{wu2019room} and PtTe$_2$($q_\mathrm{ICS}=0.003-0.033$~nm$^{-1}$)~\cite{xu2020high} but it is lower than  sputtered Bi$_x$Se$_{1-x}$ thin films~\cite{dc2018room} and comparable to Ta ($q_\mathrm{ICS}=0.1$~nm$^{-1}$).~\cite{wu2019room}

The spin degeneracy in bulk PtSe$_2$ band structure is protected due to the presence of both inversion and time-reversal symmetries.~\cite{yao2017direct} %[{\color{red} Pranaba: I changed the last sentence. Probably you would need to move the two references elsewhere, as presence of both symmetries demand the spin degeneracy. If somebody has observed the degeneracy, then a reference may be given. Else, no reference is needed.}] 
Hence, the Rashba-Edelstein effect can be ruled out as the cause of SOT.  In such cases, the origin of SOT by the interface is not very clear. % It could be some native oxide at the interface, which might induce spin splitting. But we didn't get any signature of oxygen in XPS measurements. 
Another possibility is that NiFe at the interface with PtSe$_2$ is playing a crucial role in generating SOT. In order to understand the mechanism in detail, we have performed a series of ab-initio calculations using density-functional theory (DFT). The results show that NiFe introduces three things: (a) Charge transfer from NiFe to PtSe$_2$, (b) A Rashba SOC due to the surface symmetry breaking, and (c) A proximity magnetic field that could lead to enhanced SHC in the interface region.  

    \begin{figure*}[t]
    \centering
    \includegraphics[scale = 0.7]{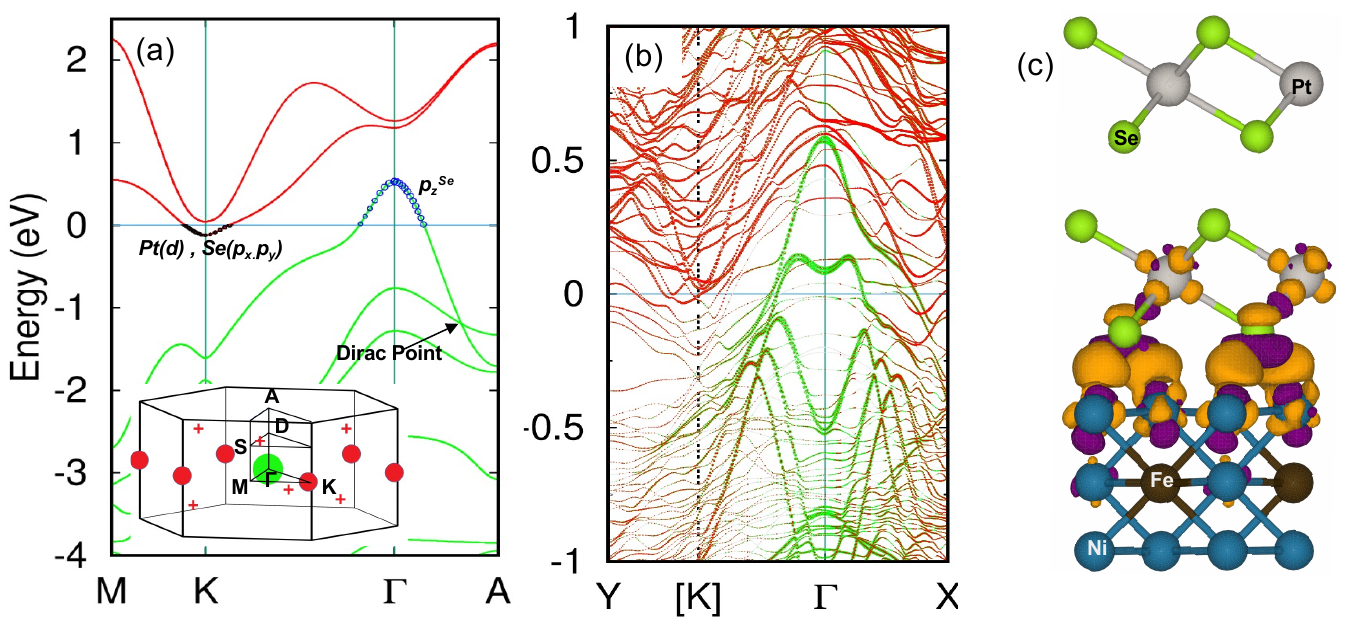}
    \caption{Density functional theory results for the PtSe$_2$/NiFe interface.
     (a) Band structure of bulk PtSe$_2$ indicating the electron pocket at K and the hole pocket at $\Gamma$ at the Fermi energy $E_F = 0$, and the type-II Dirac cone at D, along the $\Gamma-A$ line. 
     {\it Inset} shows the Brillouin zone, where the electron (green circle) and hole pockets (red circles and crosses) have been indicated.
      (b) Band structure of (PtSe$_2)_6$/NiFe showing the bulk-like bands of PtSe$_2$ (red for Pt and green for Se) with electron and hole pockets at $\Gamma$ and K, interspersed with the NiFe bands (thin lines). The BZ in PtSe$_2 /$NiFe is different from that of PtSe$_2$, but the $\Gamma$ is the same, and the projection of the K point has been indicated by the square bracket. 
      (c)  Electron charge density difference contours indicating charge transfer from NiFe to the PtSe$_2$ interface layer, due to the interface formation. The orange and purple colors indicate, respectively, the charge accumulation and depletion regions.
} 
    \label{Fig}
    \end{figure*}
Monolayer PtSe$_2$ is semi-conducting, and the gap closes for (PtSe$_2)_n$, when the number of monolayers $n \ge 4$, with the band structure tending towards that of the bulk, with electron and hole pockets as indicated in Fig. \ref{Fig} (a). In the 
(PtSe$_2)_6/$NiFe in Fig. \ref{Fig} (b), we can clearly identify the bulk-like PtSe$_2$ bands intermixed with the NiFe bands running around the gap region. 
We find a significant electron transfer to the interface PtSe$_2$ layers from NiFe ($ \sim 0.15 \  |e|$), consistent with the charge density difference contours shown in Fig. \ref{Fig} (c), suggesting an enhanced conductivity as well as spin splitting of the PtSe$_2$ band structure near the interface.

A potential mechanism for spin accumulation is the Rashba-Edelstein effect (REE)
due to the symmetry breaking at the interface.
It is difficult to compute the Rashba SOC parameter $\alpha_R$ from the PtSe$_2/$NiFe calculations due to the intermixing of the NiFe bands. 
We have instead estimated it
for the simpler case (PtSe$_2)_4 /$Ni, where
the four monolayers are enough to reproduce the bulk band structure, and a single Ni overlayer, representing NiFe, provides the symmetry breaking. 
For the top-most valence band at $\Gamma$, which represents the hole  pocket
in bulk  PtSe$_2$, we find $\alpha_R \sim 0.01 \ $eV.\AA~(see Fig.~S6 of supplementary materials),
which is rather small. However, it has been estimated that PtSe$_2$ has a rather large  momentum relaxation time $\tau \approx 80\ $fs~\cite{kurpas2021intrinsic}.
This leads to the Rashba-Edelstein length for the spin-charge conversion efficiency
to be
$\lambda_{RE} = j/j_s = \tau \alpha_R/\hbar \approx 1.2\ $ \AA. 
This relatively low efficiency would be further reduced due to the proximity magnetic field, which tends to align the spin moments out of the plane, thereby reducing the REE. 
This suggests that the REE might play some role but is not a dominant mechanism for the spin accumulation at the interface. The type II Dirac cone in bulk PtSe$_2$ is also not expected to affect spin accumulation since it occurs well below the Fermi energy.

Due to the NiFe slab, the interface PtSe$_2$ layers will experience a  proximity magnetic field, which could affect spin transport.
Even though the PtSe$_2$ layers acquire very little magnetization ($0.03 \mu_B$ for the first and $0.01 \mu_B$ for the second layer PtSe$_2$, compared to $0.59 \mu_B$ for Ni), there is a substantial spin-splitting of the PtSe$_2$ band structure due to the  presence of NiFe.
We have estimated this value to be 
 $\Delta E \approx 0.10- 0.15 $ eV
 for the PtSe$_2$ states near $E_F$ (see Fig.~S7 of supplementary materials). 
 This is significant and could lead to a large enhancement of the spin Hall conductivity (SHC).
 In fact, for the Pt$/$NiFe interface, earlier calculations by Kelly and coworkers\cite{wang2016giant} have revealed that the SHC for the Pt interface layers adjacent to the NiFe
 %within a thickness of some 20 \AA, 
 is enhanced by a factor of approximately twenty-five as compared to the Pt bulk layers. This suggests that the proximity magnetic field in the interfacial PtSe$_2$ plays an important role in the strong spin-charge conversion observed in the samples.

In conclusion, we demonstrated efficient charge-to-spin conversion in PtSe$_{2}$/NiFe/Pt using scalable large-area thin films of PtSe$_{2}$ .
Our findings indicate that the interface between PtSe$_{2}$ and NiFe can generate a significant current-induced damping-like torque. The corresponding \textit{interfacial} charge-to-spin conversion efficiency is $-0.1\pm 0.02$~nm$^{-1}$, which is three times that of the control sample NiFe/Pt as well as PtTe$_2$~\cite{xu2020high}. Our band structure calculations suggest that the proximity magnetic field in the interfacial PtSe$_2$ plays an essential role in the strong charge-to-spin conversion observed in our samples. Our work demonstrates a novel form of SOT to obtain high charge-to-spin conversion efficiencies for device applications. Additionally, our work demonstrates that "SOT through a proximity magnetic field" needs to be considered for explaining SOT observed in other TMD-based material systems.

\section{Methods}
Large-area PtSe$_{2}$ thin films were grown using thermally assisted conversion process.~\cite{jakhar2020optically} First, a 3-nm-thick Pt layer was deposited on a Si/SiO$_2$ substrate using magnetron sputtering. The base pressure of the sputter deposition system was better than $5\times10^{-8}$ Torr. The sample was then transferred to a two-zone furnace for the selenization process. The substrate with the pre-deposited Pt layer was maintained at 500~$^{\circ}$C, while the Se precursors were maintained at 280~$^{\circ}$C. N$_2$ gas was used as a carrier gas to direct Se species toward the substrate. By optimizing the substrate temperature, we found that the thermally assisted conversion of Pt into PtSe$_{2}$ takes place at a substrate temperature of 500~$^{\circ}$C.~\cite{jakhar2020optically}

Structural characterization of PtSe$_2$ was done using cross-sectional HRTEM, XRD, Raman spectroscopy, XPS and AFM. The HRTEM measurements were performed on an FEI, TECNAI G2 F30, and ST microscope with an accelerating voltage 300 kV. XRD of PtSe$_2$ film was measured in $\theta-2\theta$ mode, where X-rays of Cu-K$_\alpha$ line having a wavelength of $1.54~\mathrm{\AA}$ were used. Raman spectra of PtSe$_2$ were measured with a laser of wavelength 532 nm. XPS was recorded using AXIS Supra with the monochromatic Al-K$_\alpha$ X-ray source (1486.6 eV). The thickness of thin films was measured using AFM.

Subsequently, we fabricated microstrip ($100\times 40~\mu$m$^2$) devices for STFMR measurements using a lift-off method in two levels. In the first level, a 3-nm-thick Pt layer was deposited on the patterned area and was later selenized following the method discussed above. Using magnetron sputtering, a NiFe (5 nm)/Pt (3 nm), bilayer was subsequently deposited. In the second level, co-planar waveguides were fabricated using optical lithography and lift-off from Au(80 nm)/Cr(10 nm). 
\section*{Acknowledgement}
%\vspace{-1.5mm}
The partial support from the Science \& Engineering research board (SERB File no. CRG/2018/001012), the Ministry of Human Resource Development under the IMPRINT program (Grant no: 7519 and 7058), the Department of Electronics and Information Technology (DeitY), Joint Advanced Technology Centre at IIT Delhi, and the Grand Challenge project supported by IIT Delhi are gratefully acknowledged. Richa gratefully acknowledges the financial support from the Council of Scientific and Industrial Research (CSIR), Government of India. Theory work including the DFT calculations is funded by the Department of Science and Technology, India, through Grant No. CRG/2020/004330. SS thanks the United States-India Educational Foundation (USIEF) for their support through a Fulbright-Nehru Fellowship, jointly funded by the Governments of the United States and India. BRKN acknowledges the support of HPCE, IIT Madras for providing computational facilities.
\section*{AUTHOR DECLARATIONS}
\subsection*{Conflict of Interest}
The authors have no conflicts to disclose.
\subsection*{DATA AVAILABILITY}
The data that support the findings of this study are available from the corresponding author upon reasonable request.
%\bibliography{aipsamp}
\providecommand{\noopsort}[1]{}\providecommand{\singleletter}[1]{#1}%
\providecommand{\latin}[1]{#1}
\makeatletter
\providecommand{\doi}
  {\begingroup\let\do\@makeother\dospecials
  \catcode`\{=1 \catcode`\}=2 \doi@aux}
\providecommand{\doi@aux}[1]{\endgroup\texttt{#1}}
\makeatother
\providecommand*\mcitethebibliography{\thebibliography}
\csname @ifundefined\endcsname{endmcitethebibliography}
  {\let\endmcitethebibliography\endthebibliography}{}

\end{document}